\documentclass[final,onefignum,onetabnum]{siamonline190516}

\usepackage{lipsum}
\usepackage{amsfonts}
\usepackage{graphicx}
\usepackage{epstopdf}
\usepackage{algorithmic}
\ifpdf
  \DeclareGraphicsExtensions{.eps,.pdf,.png,.jpg}
\else
  \DeclareGraphicsExtensions{.eps}
\fi

\usepackage{enumitem}
\setlist[enumerate]{leftmargin=.5in}
\setlist[itemize]{leftmargin=.5in}

\newsiamremark{remark}{Remark}
\newsiamremark{hypothesis}{Hypothesis}
\crefname{hypothesis}{Hypothesis}{Hypotheses}
\newsiamthm{claim}{Claim}

\headers{Identifiability and Sensitivity Analysis of Kriging Weights for the Mat\'ern Kernel}{Amanda Muyskens, Benjamin W. Priest, Im\`ene R. Goumiri, and Michael D. Schneider}

\title{Identifiability and Sensitivity Analysis of Kriging Weights for the Mat\'ern Kernel }

\author{Amanda Muyskens\thanks{LLNL 
  (\email{muyskens1@llnl.gov}).}
\and 
Benjamin W. Priest \and 
Im\`ene R. Goumiri \and 
Michael D. Schneider
}
\usepackage{amsopn}

\ifpdf
\hypersetup{
  pdftitle={An Identifiability and Sensitivity Analysis of Mat\'ern Gaussian Process Hyperparameters on Prediction},
  pdfauthor={Amanda Muyskens, Benjamin W. Priest, Im\`ene Goumiri, and Michael Schneider}
}
\fi

\begin{document}

\maketitle

\begin{abstract}
	Gaussian process (GP) models are effective non-linear models for numerous scientific applications.
	However, computation of their hyperparameters can be difficult when there is a large number of training observations (n) due to the $O(n^3)$ cost of evaluating the likelihood function. 
	Furthermore, non-identifiable hyperparameter values can induce difficulty in parameter estimation.
	Because of this, maximum likelihood estimation or Bayesian calibration is sometimes omitted and the hyperparameters are estimated with prediction-based methods such as a grid search using cross validation.
	Kriging, or prediction using a Gaussian process model, amounts to a weighted mean of the data, where training data close to the prediction location as determined by the form and hyperparameters of the kernel matrix are more highly weighted.
	Our analysis focuses on examination of the commonly utilized Mat\'ern covariance function, of which the radial basis function (RBF) kernel function is the infinity limit of the smoothness parameter. 
	We first perform a collinearity analysis to motivate identifiability issues between the parameters of the Mat\'ern covariance function.
	We also demonstrate which of its parameters can be estimated using only the predictions.
	Considering the kriging weights for a fixed training data and prediction location as a function of the hyperparameters, we evaluate their sensitivities - as well as those of the predicted variance  - with respect to said hyperparameters.
	We demonstrate the smoothness parameter $\nu$ is the most sensitive parameter in determining the kriging weights, particularly when the nugget parameter is small, indicating this is the most important parameter to estimate.
	This is opposite of GP fitting convention, where $\nu$ is fixed. 
	Finally, we demonstrate the impact of our conclusions on performance and accuracy in a classification problem using a latent Gaussian process model with the hyperparameters selected via a grid search. 
\end{abstract}


\section{Introduction}
\label{sec:intro}
Gaussian process models are popular nonlinear models for prediction in both the fields of machine learning and statistics. 
GPs are continuous generalizations of the multivariate normal distribution, where any subset of data sampled over a specified domain is assumed to be jointly multivariate normally distributed with specified mean and positive semi-definite covariance (alternatively: kernel) functions.
Unlike simple regression that assumes independent observations, Gaussian process regression (GPR) accounts for correlation among observations. 
GPs are appropriate models for data collected over space, time, or as functional approximators for highly non-linear responses.  
The primary goal of GPR is the estimation of a response at fixed new interpolated locations, which is given by a predicted posterior mean. 
These models can be trained with a small number of observations, and unlike many other machine learning methods, uncertainty quantification can be directly estimated through explicit conditional distributions.

In order to guarantee a valid model, one typically assumes a parametric form for the covariance matrix, usually called a kernel matrix.
The hyperparameters guiding that covariance are often considered nuisance hyperparameters because their estimation is often difficult and computationally expensive. 
When there are a large number of observations, realizing the kernel matrix for the purposes of likelihood evaluation can become prohibitive in terms of memory and time. 
Furthermore, hyperparameter optimization can itself be difficult since the estimation surface can be multi-modal and some covariance forms can exhibit non-identifiability among their parameters.
A grid search using cross validation is a pragmatic alternative hyperparameter estimation method commonly used in the machine learning literature \cite{rasmussen2006gaussian}, but is also expensive for large or high-dimensional data as it requires repeated matrix inversion of the large covariance matrix. 
Further, the number of evaluations grows exponentially with the number of parameters to be simultaneously estimated. 
MuyGPs is a modern prediction-based alternative training method that emphasizes computational efficiency  \cite{muyskens2021muygps}.
Instead of using maximum likelihood estimation, this method estimates kernel hyperparameters only based only on leave-one-out prediction accuracy using local kriging estimators.
However, it is not obvious whether prediction-based methods are sufficient in order to estimate the hyperparameters as needed for Gaussian process regression.


Variance-based global sensitivity analysis is a useful tool in understanding non-linear parameter relationships in numerous applications. 
This type of analysis evaluates the proportion of variance in a response to corresponding changes in the input parameters of interest. 
The analysis determines a parameter's sensitivity to be proportional to its variance.
This information can then be used to restrict the number of estimated parameters, where insensitive parameters can be fixed at nominal values.
Accordingly, sensitivity analysis provides guidance for selecting which among a set of unidentifiable parameters to fix.
Sensitivity analysis has been applied to Gaussian processes through several unique analyses. 
\cite{chen2018priors} determine that the sensitivity of hyperparameter priors on their posteriors have no significant effect. 
\cite{paananen2019variable} use sensitivity analysis to show that downselecting the included training locations is more performant than using the nearest neighbors. 
\cite{stephenson2021measuring} demonstrate a pipeline to evaluate the sensitivity of inference from GPR with respect to the kernel selection.
Finally, \cite{svenson2014estimating} use Gaussian process models as surrogate models in order to avoid expensive computer model evaluations.

However, these studies fail to capture the relationship between kernel hyperparameter estimates on predictions.
The predictions from very different kernels applied to a highly correlated dataset may be correlated due to the data correlation itself rather than any similarity in the prediction procedures.
Therefore, to decouple the data dependence, we explore the sensitivity of parameter values for one of the most popular covariance functions, the Mat\'ern covariance.
We explore the the functional sensitivity of values on the kriging weights, which are the weights applied to the data in order to perform kriging, and the kriging conditional variance.
In Section~\ref{background}, we describe the mathematical details of a Gaussian process and its estimation. 
In Section~\ref{ident}, we explore the identifiability of the parameters of the Mat\'ern kernel function as a function of predictions.
In Section~\ref{sensitivity}, we generally describe sensitivity analysis and the BASS emulator. 
In Section~\ref{results}, we describe the training, emulation, and analysis involved in our sensitivity analysis. 
Section~\ref{data} demonstrates the practical performance of the results of our sensitivity results in the classification of star and galaxy images using latent GP classification. 
Finally, in Section \ref{sec:conclusions} we summarize our contributions and discuss future work.

\section{Background}
\label{background}


In the most general sense, a Gaussian process is the generalization of the multivariate normal distribution to a continuous domain.
 Assume we observe a univariate response $n\times 1$ vector $Y$ that is assumed to follow a Gaussian process. For simplicity and as is conventional, we assume $Y$ is de-trended and therefore mean zero, but the extension of this to non-zero processes is trivial. We observe this response at $n$ parameter $p \times 1$ vector values $x=(x_1^T, x_2^T, ...,x_n^T)^T$. Let $Y$ be the $n \times 1$ vector of observations observed at those parameters so $Y=(Y(x_1),Y(x_2),...,Y(x_n) )^T$.  Then, the Gaussian process implies every finite combination of observations are jointly normal so that
$$Y(x) \sim MVN\Big(\widetilde{0}, K_\theta(x,x)\Big),$$
where $\widetilde{0}$  is the $n\times 1$ vector is populated with zeros, and $K_\theta(x,x)$ is a $n \times n$ positive definite, symmetric covariance matrix that is controlled non-linearly through parameters $\theta$.
Similarly, any unobserved parameter set $x^{\star}$ is also assumed jointly normal with our observed data as
\begin{eqnarray*}
	\begin{pmatrix}Y(x)\\
		Y(x^{\star})
	\end{pmatrix} & \sim & MVN\left[\left(\begin{array}{c}
		\widetilde{0}\\
		0
	\end{array}\right),\left(\begin{array}{cc}
		K_\theta(x,x) & K_\theta(x,x^{\star})\\
		K_\theta(x^{\star},x) & K_\theta(x^{\star},x^{\star})
	\end{array}\right)\right].\\
\end{eqnarray*}

Then, the conditional distribution for the response at the new parameter set $x^{\star}$ is also multivariate normal with mean and variance
\begin{align}
\widehat{Y}(x^{\star}|x) 
&= K_\theta(x^{\star}, x) K_\theta(x,x)^{-1} Y(x) \textrm{, and}
\label{predmean} \\
\text{Var}(\widehat{Y}(x^{\star}|x))  
&= K_\theta(x^{\star}, x^{\star}) - K_\theta(x^{\star}, x) K_\theta(x,x)^{-1} K_\theta(x, x^{\star}).
\label{predvar}
\end{align}

Note that the conditional mean is the best linear unbiased predictor (BLUP) even when the normality assumption of the Gaussian process is violated. 
$K_\theta(x^{\star}, x) K_\theta(x,x)^{-1}$ is a $1 \times n$ vector of weights that are applied to the de-trended data vector $Y$ used to compute its predicted value. 
For the remainder of this manuscript we refer to this vector as the kriging weights, since it is a weighting applied to the data vector to perform kriging.
For the Mat\'ern covariance, there is no general closed form explicit function to directly compute these weights without forming and inverting the matrix $K_\theta(x,x)$, which can be prohibitively expensive in large training data. 
We refer to Equation \ref{predvar} as the kriging variance.

Then the log-likelihood of the observations $Y$ follow a jointly multivariate normal distribution:
\begin{equation}
\label{ll}
log(L(\theta, Y)) = - \frac{p}{2}log(2 \pi)  - \frac{1}{2} log(|K_\theta(x,x)|)  - \frac{1}{2} Y^T K_\theta(x,x)^{-1} Y,
\end{equation}
where $p$ is the dimension of the vector $x_i$ for $i=1,2,...n$. 
Often $K_\theta(x,x)$ is assumed to be stationary and isotropic. That means that the covariance function is a function of only the distances between the parameters. For parameter vectors $x_1$ and $x_2$
\begin{equation}
K_\theta (x_1, x_2) = \phi_\theta(||x_1 - x_2||_2),
\end{equation}
where $\phi_\theta$ is assumed to be a functional form with parameters $\theta$. 
One of the most common covariance forms $\phi_\theta$ is the  Mat\'ern covariance function. With covariance hyperparameters $\theta = \{\sigma^2, \rho, \nu, \tau^2\}$, and where $d=||x_i - x_j||_2$ is the distance between two locations $x_i, x_j$ in $x$.
\begin{equation}
\phi_\theta(d) = \sigma^2 \frac{2^{1- \nu}}{\Gamma(\nu)} \Bigg( \sqrt{2\nu} \frac{d}{\rho} \Bigg)^{\nu} K_\nu \Bigg( \sqrt{2\nu} \frac{d}{\rho} \Bigg) + \tau^2 \mathbb{I} \{d=0\},
\end{equation}
where $K_\nu$ is the modified Bessel function of the second kind. 
As $\nu \to \infty$, this form converges pointwise to the Gaussian (RBF) covariance function.  
Although this kernel form is complex and contains a Bessel function, when $\nu$ is $\{\frac{1}{2},\frac{3}{2},\frac{5}{2},...\}$, it takes on relatively simple forms.  For example, when $\nu = \frac{1}{2}$, the form simplifies to the Exponential covariance function 

\begin{equation}
\phi_\theta(d) =\sigma^2 \exp{-\frac{d}{\rho}} + \tau^2 \mathbb{I} \{d=0\}.
\end{equation}
These forms are simpler and faster to evaluate. 
Therefore, it is common practice to assume one of these fixed $\nu$ values \cite{gu2018robustgasp}.
Further it is common practice to fix $\tau^2$ at a nominal value in lieu of its estimation. 
Particularly, $\tau$ can be assumed 0 in deterministic computer experiments, or it is often assumed to be a small value such as 0.001 in order to aide in numerical stability in matrix inversion.

Using the log-likelihood in Equation \ref{ll}, the covariance parameters $\theta$ can be estimated via maximum likelihood estimation, Bayesian calibration using Markov Chain Monte Carlo (MCMC), by prediction-based method like grid search cross validation, or some other estimation method.
We focus on prediction-based methods, and evaluate the estimability of the Mat\'ern hyperparameters through their identifiability and sensitivities.
Despite the benefits of flexility of the Mat\'ern kernel, there are known difficulties in estimating the parameters of the Mat\'ern covariance, where similar correlation functions can be obtained from non-unique sets of parameters. 
Namely, an increased smoothness parameter $\nu$ can be compensated with a smaller estimate of the range parameter $\rho$ to create a similar covariance function and vice versa. 
Further, hyperparameter estimation can be difficult because the surfaces are non-convex, and the likelihood is computationally expensive to compute ($O(n^3)$). 
We aim to determine best-practice for estimation of these hyperparameters in parameter estimation based on predictions such as a grid search or MuyGPs \cite{muyskens2021muygps}.

\section{Identifiability of Mat\'ern hyperparameters}
\label{ident}

Define $\Omega_\theta$ to be the function that is a scaled version of the kernel function $K_\theta$ such that $\Omega_\theta(x,x) = \frac{K_{\theta}(x,x)}{\sigma^2} $. 

Then, we can write the kriging weights vector $Z$ as

$$Z=\sigma^2\Omega_{\theta}(x^{\star}, x) (\sigma^2\Omega_{\theta}(x, x) + \tau^2I)^{-1}.$$
Then, if we define $\omega^2 = \frac{\tau^2}{\sigma^2}$, we can re-write $Z$ as
$$Z=\Omega_{\theta}(x^{\star}, x) (\Omega_{\theta}(x, x) + \omega^2I)^{-1}$$
This demonstrates predictions themselves only depend on the parameters $\sigma^2$ and $\tau^2$ though their ratio $\omega^2$. 
From this, we conclude that using the traditional parameterization, one cannot uniquely estimate both $\sigma^2$ and $\tau^2$ using cross validation or any other estimation method based on predictions only. 
Note that this is the case regardless of the functional form of the covariance matrix, and is true in other forms in general. 
For this reason, we adopt this $\omega^2$ parameterization and evaluate the sensitivity of the reduced set $\theta = \{\rho, \nu, \omega^2\}$. 

In contrast, the $\omega^2$ parameterization is insufficient in the prediction covariance, which does depend explicitly on both $\sigma^2$ and $\omega^2$.

$$\text{Var}(\widehat{Y}(x^{\star}|x)) = \sigma^2 \Omega(x^{\star},x^{\star}) - \sigma^2 \Omega_{\theta}(x^{\star}, x) (\Omega_{\theta}(x, x) + \omega^2I)^{-1}\Omega_{\theta}(x, x^{\star}).$$

\begin{figure}
	\centering
	\includegraphics[scale=.55]{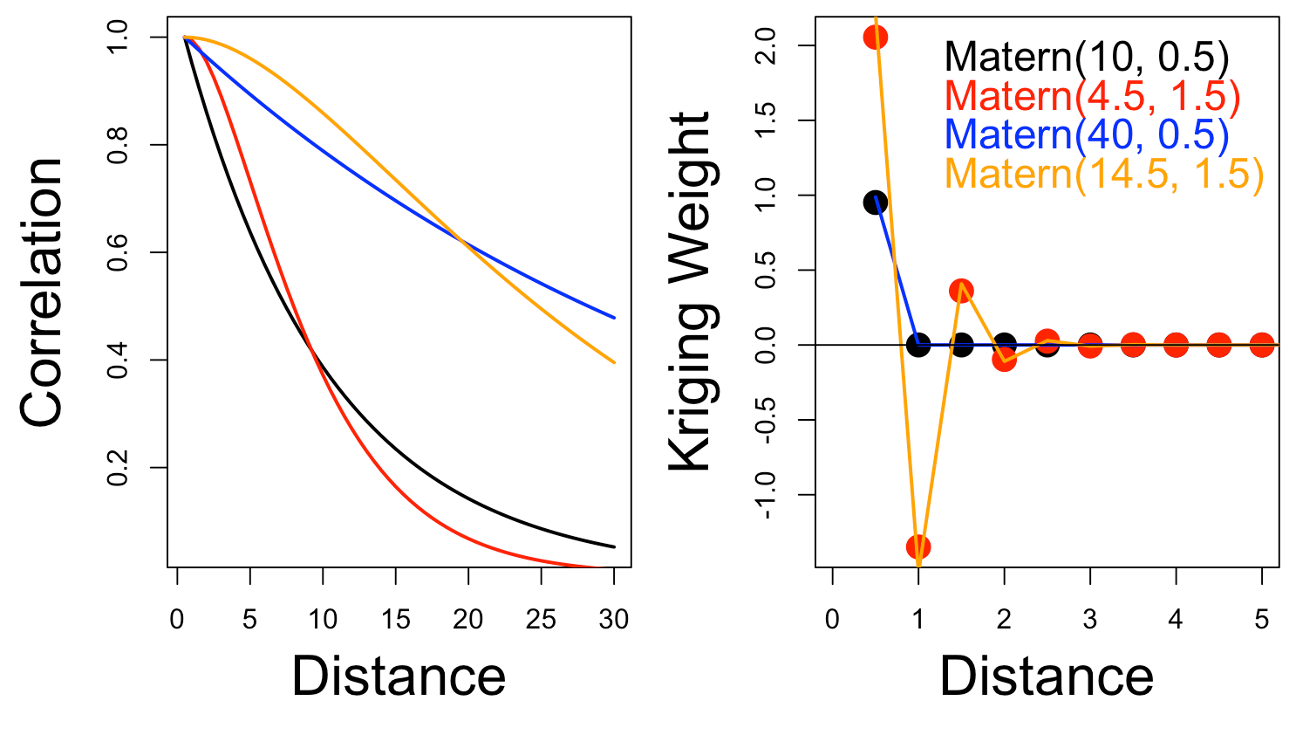}
	\caption{Comparison of similarity in correlation function and kriging weights for several $(\rho, \nu)$ values using the Mat\'ern kernel.}
	\label{fig:ck}
\end{figure}

Further, we explore other potential identifiability concerns between parameters $\nu$ and $\rho$. 
In Figure \ref{fig:ck}, we demonstrate that very different correlation functions can produce similar kriging functions and vice-versa. 
Therefore we explore identifiability in terms of both the correlation function, as well as the kriging weights.
In order to investigate these, we perform a formal global identifiability analyses via evaluation of the collinearity of $\nu, \rho$ for the correlation and kriging weights functions.

Define $S$ to be the matrix of the local sensitivities for each parameter for each function.
Then the corresponding collinearity index is defined as
\begin{equation}
\gamma =\frac{1}{\sqrt{\min(EV(S^TS))}},
\end{equation}
where $EV(*)$ is a function that extracts the eigenvalues of the argument \cite{brun2001practical}.
The collinearity index's exceeding a fixed cutoff threshold (usually between 10-20) indicates highly collinear variables, and therefore detects identifiability issues.

We sample 10,000 parameters on a grid for $\nu$ in the range of (0.01, 2.5), and $\rho$ in (0.01, 5.0), and report the corresponding correlation and kriging weight functions.
For illustration, we consider $x$ to be a length 20 one-dimensional grid where $x\in[0,1]$, excluding the prediction location $x^{\star}=0.5$.
The collinearity is computed using the \textbf{FME} package in the computing language R using the function ``collin'' \cite{soetaert2010inverse}. 

\begin{figure}
	\centering
	\includegraphics[scale=.55]{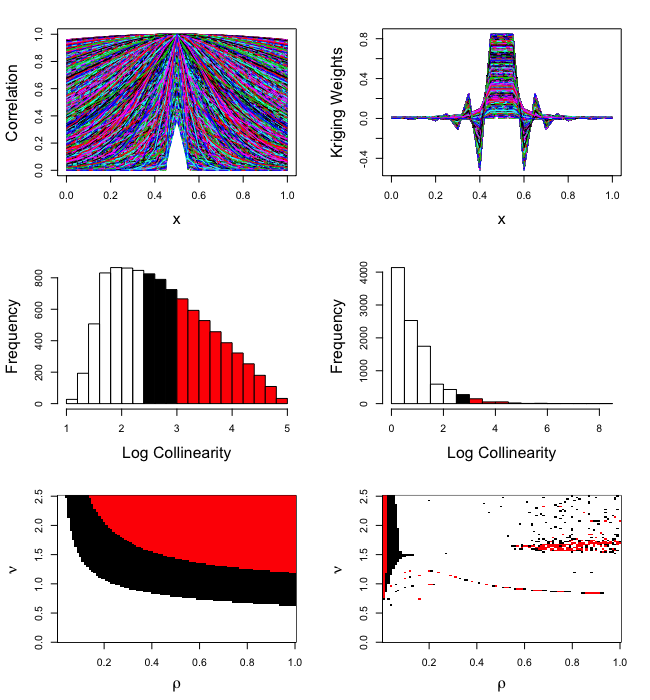}
	\caption{Collinearity of Mat\'ern hyperparameters with respect to the correlation function (left colum) and kriging weights (right column). 
		The first row demonstrates the training functional outputs. The second row are histograms of local collinearity across the hyperparameter space. The third row is a plot of local collinearity plotted by the hyperparameter values. White are locally identifiable parameter sets with collinearity $<10$, black may be identifiable with $10<$collinerity$<20$, and red are highly collinear with collinearity$>20$.}
	\label{fig:ident}
\end{figure}

Results from these analyses are in Figure \ref{fig:ident}.
In the correlation function, there are obvious bands where the increased smoothness the function demonstrates more collinearity between the parameters. 
This makes sense since as $\nu \to \infty$, the Mat\'ern kernel converges pointwise to the RBF kernel that has one less parameter than the Mat\'ern kernel. 
However, the pattern of global identifiability is much less well defined when considering the kriging weights. 
In both cases, we can see that the Mat\'ern hyperparameters $\nu$ and $\rho$ are collinear in multiple locations across their domains, and therefore are not globally identifiable with respect to the correlation function or the kriging weights individually.
Therefore, it is important to study the sensitivity of their impact on prediction so that the most appropriate hyperparameter can be fixed in estimation.

\section{Sensitivity Analysis Using Bayesian Adaptive Smoothing Splines (BASS)}
\label{sensitivity}

Global sensitivity analysis apportions the variation of a response to the variation in the respective changes in input variables. 
In over-parameterized functions, it evaluates and compares the contributions of unidentifiable parameters to the response.
It then fixes the least sensitive variables at nominal values in order to uniquely estimate the remaining subset of sensitive parameters \cite{smith2013uncertainty}.
Define $Z_\theta$ to be the random variable from $\mathbb{R}^3 \to \mathbb{R}^n$ that takes in the hyperparameters $\theta = \{\rho, \nu, \omega^2\}$ and returns the vector kriging weights at fixed locations. 
\begin{equation}
Z_\theta = K_\theta(x^{\star}, x) K_\theta(x,x)^{-1}
\end{equation}

We can decompose $Z$ into its conditional components so that 

$Z _\theta=  $
$$E(Z_\theta) +  E(Z_\theta|\rho)  + E(Z_\theta|\nu) + E(Z_\theta|\omega^2) +
E(Z_\theta|\rho,\nu)+ E(Z_\theta|\nu, \omega^2)+ E(Z_\theta|\rho,\omega^2)+ E(Z_\theta|\rho,\nu, \omega^2).$$
Since these components are independent by construction, the variance of $Z_\theta$ can be written as

$$\text{Var}(Z_\theta)
=  V_{\rho} + V_{\nu} +V_{\omega^2} +  V_{\rho,\nu} + V_{\nu, \omega^2} + V_{\rho, \omega^2} + V_{\rho,\nu, \omega^2}.$$

A common measure of sensitivity is the total variance, which is defined as the sum of these variance and the higher-order interaction terms that include the parameter of interest. 
For example, the total variance for the parameter $\rho$ is
 $$V_{\rho}  + V_{\rho,\nu} + V_{\rho, \omega^2} + V_{\rho,\nu, \omega^2}.$$

There are numerous algorithms to estimate these global sensitivity measures, particularly in the scalar response case \cite{saltelli1999sensitivity}. 
However, for fixed training data locations $Z_\theta$ is a vector rather than a univariate response. 
Although these are ultimately discrete points rather than a continuous function, for evaluation of their sensitivities, we consider the kriging weights a functional response for evaluation purpose.
In this case, computing the global sensitivities is more challenging.
One could consider the sensitivity of each kriging weight individually, but since the functional observations are correlated, it is more ideal to consider their variance modeled simultaneously. 
\cite{francom2020bass} notes that if the functional response is emulated using the Bayesian multivariate adaptive regression splines (BMARS), the sensitivity of the functional response is easily estimable. 
This method constructs a model for the vector $Z_\theta$ for evaluation at general $\theta$. 
Formally,
$$Z_\theta = f(\theta) + e,$$
where $e \sim N(0,\sigma^2)$ and $f(\theta)$ is comprised of splines so that $f(\theta) = a_0 + \sum_{m=1}^{M}a_mB_m(\theta)$, and $B_m(\theta)$ is a tensor product of spline basis functions referred to as BASS basis functions.
These basis function components are composed so that there are closed-form estimates of the sensitivities of this numerical twin.
Further details on the the computation of these variance components in this model can be seen in \cite{francom2020bass}.
We report and compare the total global sensitivies of the kriging weights as a function of kernel hyperparameters $\theta$ for fixed $x$ in the next section.

\section{Numerical Results}
\label{results}
In this section, we describe the specifics of several sensitivity studies conducted. 
Using randomized Latin hypercube designs, we sampled 5,100 hyperparameter sets for the Mat\'ern covariance within the bounds in Table \ref{table:1}. 
Note that $\sigma^2$ is not utilized in sensitivities of kriging weights since it is non-identifiable, but is included in the sensitivities of the kriging variance.
These ranges are meant to represent ranges that  for covariance functions on a domain on the $p$ dimensional unit hypercube $[0,1]^p$, and normalized for scale and center.
5,000 of these observations are used to train BMARS emulator using the \textbf{BASS} package in the programming language R \cite{francom2020bass} and 100 observations are reserved for out of sample emulator testing. 
Note that the sensitivities reported are those of the emulator, not the data itself. 
Therefore, if the emulator represents an adequate likeness to the data, the results can be interpreted as data sensitivities, but this likeness must be evaluated for fidelity of the inference.
 
Unlike a correlation function that can easily be realized for all theoretical distances between points, the kriging weights themselves are more complex to study since they are dependent on the choice of observed $x$ and the prediction location $x^{\star}$. 
For simplicity, we first assume a gridded $x$ structure surrounding the prediction location $x^{\star}$ in low dimensional cases of 1 or 2 dimensional space.
These low-dimensional domains are common in data collected at fixed locations either over time (1-D) or space (2-D).
Further, we must also select a density of the gridded locations $x$.
Figure 2 demonstrates kriging weights for the same set of parameters sampled at various grid densities. 
Very similar kriging weights are given regardless of grid density for the closest, and therefore most important observations.
Therefore, we evaluate 20 training locations in the 1-D case, and 16 in the 2-D case, but expect these results to generalize to other grid densities and possibly dimensions. 
The results for these sensitivity analyses can be seen in Tables \ref{table:2} and \ref{table:3} .

\begin{table}[h]
	\centering
	\begin{tabular}{c c c} 
		Parameter & Minimum & Maximum \\ [0.5ex] 
		\hline
		$\sigma^2$ &  0.1 & 5 \\ 
		$\rho$ & 0.01 & 5 \\
		$\nu$ & 0.01 & 2.5  \\
		$\omega^2$ & 0.001 & 0.1
	\end{tabular}
	\caption{Parameter ranges involved in the global sensitivity analysis.}
	\label{table:1}
\end{table}

\begin{figure}
	\centering
	\includegraphics[scale=.4]{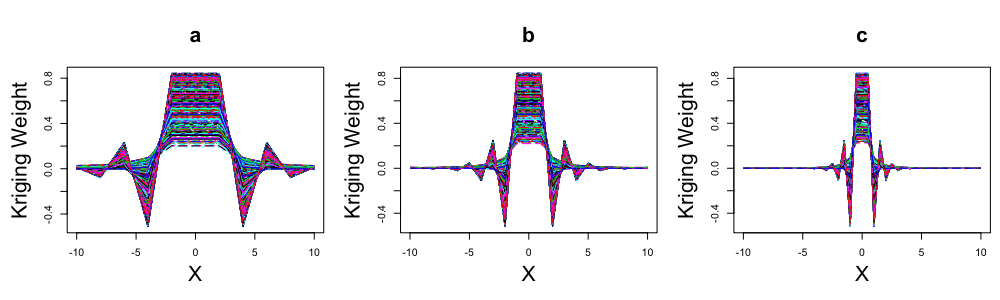}
	\caption{Mat\'ern kriging weights for grids with intervals of 2 (a), 1 (b), and 0.5 (c).}
\end{figure}

\begin{table}[h]
	\centering
	\begin{tabular}{r|rrrrr}
	& $\sigma^2$ & $\rho$ & $\nu$ & $\omega^2$ & x \\ 
	\hline
	weights $\omega^2=0$ & NA & 0.1 & 22.4 & NA & 77.6  \\ 
	weights $\omega^2=0.001$ & NA & 18.5 & 29.8 & NA & 51.7\\ 
	weights $\omega^2=0.01$ & NA & 14.5 & 31.3 & NA & 54.2 \\ 
	weights $\omega^2=0.1$ & NA & 17.9 & 22.4 &  NA & 59.6\\ 
	weights $\omega^2$ vary & NA &18.7& 26.0 & 11.8& 43.5 \\ 
	Prediction variance & 25.7& 10.4 & 60.3 & 3.6 & NA \\ 
\end{tabular}
	\caption{Percent of total Sensitivities for 1-D data. Each row repesents a different BASS emulator, and each row is normalized to be a percent for comparability across settings.}
	\label{table:2}
\end{table}

\begin{table}[h]
	\centering
	\begin{tabular}{r|rrrrr}
		& $\sigma^2$ & $\rho$ & $\nu$ & $\omega^2$ & x \\ 
		\hline
		weights $\omega^2=0$ & NA & 0.3 & 14.3 & NA & 85.4  \\ 
		weights $\omega^2=0.001$ & NA & 11.6 & 18.8 & NA & 69.6 \\ 
		weights $\omega^2=0.01$ & NA & 12.9 & 19.1 & NA & 68.0 \\ 
		weights $\omega^2=0.1$ & NA & 15.2 & 19.8 &  NA & 65.0\\ 
		weights $\omega^2$ vary & NA &17.1& 20.9 & 9.8& 52.2 \\ 
		Prediction variance & 24.9& 20.4 & 52.5 & 2.3 & NA \\ 
	\end{tabular}
	\caption{Percent of total Sensitivities for 2-D data. Each row repesents a different BASS emulator, and each row is normalized to be a percent for comparability across settings.}
	\label{table:3}
\end{table}

\section{Star-Galaxy classification}
\label{data}

To demonstrate the usefulness of our sensitivity analysis findings, we compare predictions from estimation of combinations of the Mat\'ern hyperparameters.  
The general field of image classification is dominated by machine learning, particularly convolutional neural networks \cite{lu2007survey}. 
However, in \cite{muyskens2021star}, it is demonstrated that classification via Gaussian processes can be competitive, particularly  in small training data size, when model reduction is performed prior to modeling on the classification of images as either stars or galaxies.
This application is vital to parsing large space surveys such as Rubin Observatory Legacy Survey of Space and Time (LSST) that will have immense data that will need to be automatically labeled for reconstruction of the mass-density of the universe.
One example of the utility of this type of analysis is to compare the state of the universe to simulations in order to develop posteriors for dark matter parameters \cite{schneider2017probabilistic}.
Here, we we utilize this application as a benchmark example for sensitivity of predictions, and in this study, we will not make use of the uncertainty quantification given from the latent GP models as in \cite{muyskens2021star}. 

\begin{figure}
	\centering
	\includegraphics[scale=.45]{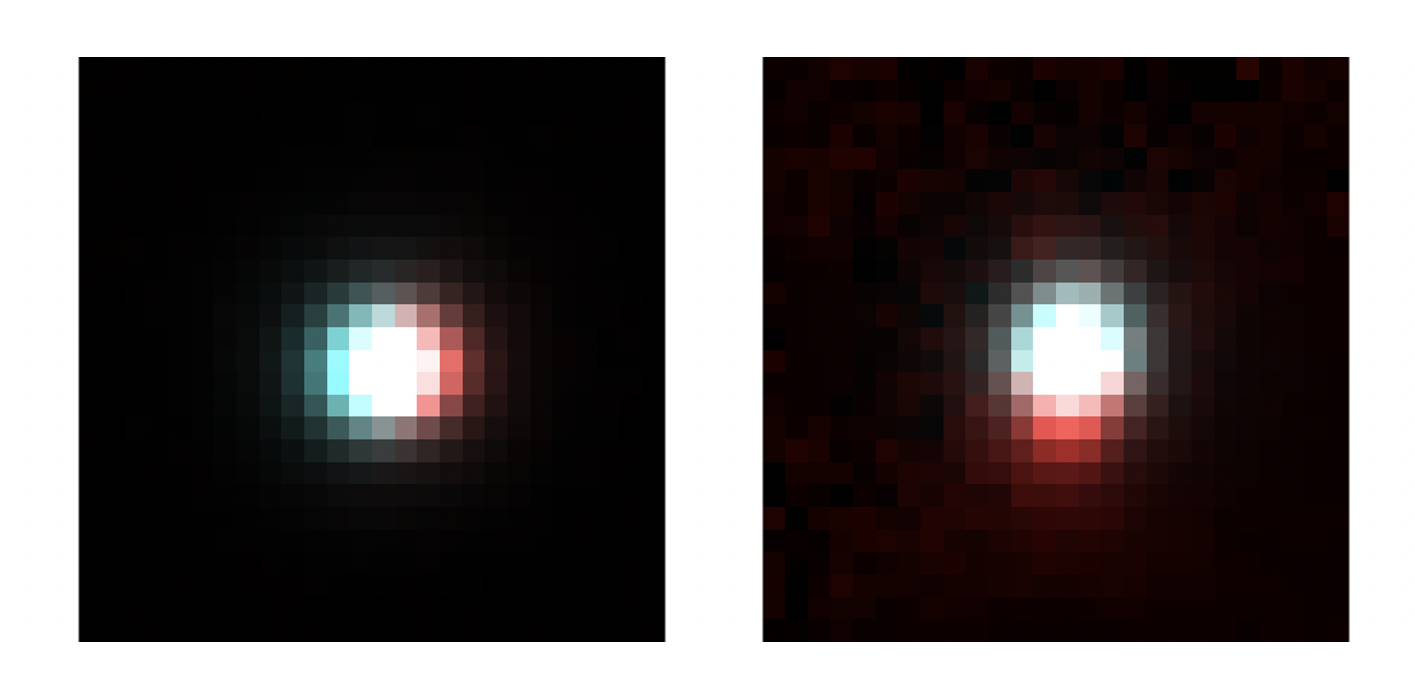}
	\caption{Example images of a star (left) and a galaxy (right), where the red coloring is image data from the $i$ band, and blue is the corresponding point spread function (PSF). }
	\label{fig:stargal}
\end{figure}

Classification via Gaussian Processes involves defining a latent variable on which the Gaussian process is fit. 
Define $Y(x)$ as a latent Gaussian process, with the Gaussian process description in Section \ref{background}. 
Images classified as a star are designated 1, and those known to be a galaxy are alternatively designated -1.  
An approximate principal components analysis to is applied to images in order to embed them into a lower-dimensional space prior to inference as in \cite{goumiri2020star}.
Then classification is given as 
 \begin{equation}
  Q(x) =
 \begin{cases}
 1 \  \ \ (\text{star}) &  \text{if Y(x)} > 0\\
 -1 \  (\text{galaxy}) & \text{if Y(x)} < 0
 \end{cases}.
 \end{equation}

We compare average prediction accuracy and average computing time in 50 random simulation iterations using the latent Gaussian process classification model above. 
Once the data is cleaned, we have a total of 8 images of each object, where 4 there are different band images and point spread functions (psf) with a total of 23780 pixels, which we refer to as an image set. These image sets are normalized to have norm 1. 
In each iteration, 500 test image sets are randomly selected where 250 image sets are stars and 250 image sets are galaxies, and they are reserved for testing. 
From the remaining available data, we select testing data equally in star and galaxy labeling for various training set image sizes.
As the sample size increases within an iteration, we retain the lower size training samples, and only add new data for consistency.
We compute the first 30 approximate eigenvectors using the ``partial\_eigen'' function from the R package ``irlba'' and multiply them with the original data matrix to create a reduced dataset on which we perform the classification. 
For hyperparameter estimation, we perform a grid search using leave-one-out cross validation, where we seek to maximize the classification accuracy. 
If multiple parameter sets yield the same training accuracy, the mean hyperparameter estimate is selected.
Instead of doing full prediction that involves all the data points, we produce local predictions where we use the closest 50 nearest neighbors \cite{muyskens2021star,}. 
In each parameter, we consider 10 gridded values as described in Table \ref{table:1}.

\begin{figure}
	\centering
	\includegraphics[scale=.45]{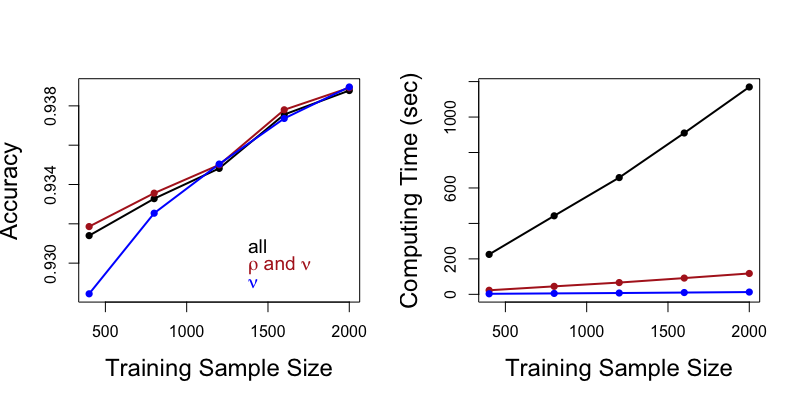}
	\caption{Mean accuracy (left) and mean computing time (right) for grid search hyperparameter estimation and prediction using subsets of the Mat\'ern hyperparameters.}
	\label{fig:dat}
\end{figure}

Results for the accuracy and computing time are in Figure \ref{fig:dat}. 
In estimation for a smaller number of training images $(n=400, 800)$, the mean accuracy of only estimating the $\nu$ parameter is lower than that of estimating all parameters. 
However, for the training image sizes $n=1200, 1600, 2000$, there is no difference in accuracy performance among the methods. 
However, in terms of computing time, there is a very large improvement speedup by fixing more parameters. 
By estimating only $\nu$, when $n=2000$, the entire analysis takes on average 12.8 seconds. 
That is a roughly 91x speed up when compared to the equally performant methods of estimating all parameters and a 9.2x speed up compared to estimating the two parameters $\nu$ and $\rho$. 
This speed up is due to the fact we compute predictions for only 10 parameter sets in the $\nu$ only case, but 1,000 sets in the all parameters estimation case. 
Therefore we conclude that in this dataset for larger sample sizes, it is preferable to do a grid search in only the parameter $\nu$. 
This agrees with the sensitivity analysis so that $\nu$ is the most important parameter to estimate correctly. 
Ultimately, this allows for equally accurate performance in large sample sizes with only estimation of one hyperparameter, which allows for fewer evaluations, especially in repeated research models where uncertainty quantification is not needed.

\section{Conclusions}
\label{sec:conclusions}

In this manuscript, we have performed an identifiability and sensitivity analysis of the kriging weights and kriging variance using the Mat\'ern covariance. 
When there is no measurement error ($\tau^2=0$), the prediction weights are determined almost entirely by the smoothness parameter $\nu$, as well as the location of the observation. 
However, as $\tau^2$ increases, the hyperparameter $\rho$ becomes increasingly sensitive, particularly for large estimates of $\nu$. 
The kriging variance is sensitive to all Mat\'ern hyperparametetrs. 

In terms of efficient hyperparameter estimation practice, these results lead us to several conclusions. 
First, if cross validation is to be used for parameter selection, $\tau^2$ and $\sigma^2$ are non-identifiable as traditionally parameterized. 
However, we can define a new parameter $\omega^2 = \frac{\tau^2}{\sigma^2}$ that can be estimated by this method, but this limits the parameters to $\rho, \nu, \omega^2$.
Next, in any type of hyperparameter estimation, if a hyperparameter is to be fixed to assist with the identifyability issues in estimation, the range parameter should instead be fixed rather than current practice where the smoothness parameter is fixed.  
Finally, we demonstrated the practical implications of these sensitivity results for the classification of stars and galaxies using a latent Gaussian process. 
Using a grid search only in the smoothness parameter $\nu$, we were able to speed up estimation by approximately 91x with no change in kriging accuracy.
Future work could perform a similar analysis on other kernel forms such as the periodic covariance function or dual kernels that are limiting cases of neural network architechtures such as the neural tangent (NTF) kernel 
\cite{jacot2018neural}.

\appendix
\section{An example appendix}

\section*{Acknowledgments}
This work was performed under the auspices of the U.S. Department of Energy by Lawrence Livermore National Laboratory under Contract DE-AC52-07NA27344 with IM release number LLNL-JRNL-827417. Funding for this work was provided by LLNL Laboratory Directed Research and Development grant 19-SI-004.
This document was prepared as an account of work sponsored by an agency of the United States government. Neither the United States government nor Lawrence Livermore National Security, LLC, nor any of their employees makes any warranty, expressed or implied, or assumes any legal liability or responsibility for the accuracy, completeness, or usefulness of any information, apparatus, product, or process disclosed, or represents that its use would not infringe privately owned rights. Reference herein to any specific commercial product, process, or service by trade name, trademark, manufacturer, or otherwise does not necessarily constitute or imply its endorsement, recommendation, or favoring by the United States government or Lawrence Livermore National Security, LLC. The views and opinions of authors expressed herein do not necessarily state or reflect those of the United States government or Lawrence Livermore National Security, LLC, and shall not be used for advertising or product endorsement purposes.

\bibliographystyle{siamplain}
\bibliography{sens}

\end{document}